\documentclass[twocolumn,aps,prl,superscriptaddress,showpacs,floatfix]{revtex4-1}
\usepackage{graphicx}
\usepackage{amsmath}
\usepackage{amssymb}
\usepackage{hyperref}
\usepackage{bm}
\usepackage{color}

\begin{document}

\title{Enhancing quantum order with fermions by increasing species degeneracy}
 
\author{Khadijeh Najafi}
\affiliation{Department of Physics, Georgetown University, Washington, D.C. 20057-0995, U.S.A.}
\author{M. M. Ma\'ska}
\affiliation{Department of Theoretical Physics, University of Silesia, Katowice, Poland}
\author{Kahlil Dixon}
\affiliation{Department of Physics \& Astronomy, Louisiana State University, 202 Nicholson Hall, Baton Rouge, Louisiana, 70803, U.S.A.}
\author{P. S. Julienne}
\affiliation{Joint Quantum Institute, University of Maryland and National Institute for Standards and Technology, College Park, Maryland 20742, U.S.A.}
\author{J. K. Freericks}
\affiliation{Department of Physics, Georgetown University, Washington, D.C. 20057-0995, U.S.A.}
\date{\today}

\begin{abstract}
One of the challenges for fermionic cold atom experiments in optical lattices is to cool the systems to low enough temperature that they
can form quantum degenerate ordered phases. In particular, there has been significant work in trying to find the antiferromagnetic phase transition of the Hubbard model in three dimensions, without success.  Here, we attack this problem from a different angle by enhancing the ordering temperature via an increase in the degeneracy of the atomic species trapped in the optical lattice. In addition to developing the general theory, we also discuss some potential systems where one might be able to achieve these results experimentally.
\end{abstract}
\maketitle
While the off-diagonal long-range order in cold bosonic atomic gases has been 
observed many years ago, quantum magnetism in fermionic gases is still a challenge
for experimentalists. Despite the possibility to control the interaction between spin
states of atoms in an optical lattice \cite{duan2003}, the temperatures
required to obtain magnetic ordering remain lower than those achievable with
current techniques. 
Therefore, it is 
easier to demonstrate the presence of magnetic correlations, before the true long-range
magnetic order is established. Using the spin-sensitive Bragg scattering of light,
antiferromagnetic correlations in a two-spin-component Fermi gas, magnetic correlations have been observed
at a temperature 40\% higher than the putative temperature for the transition to the 
antiferromagnetic state in three dimensions \cite{hart2015}. In this experiment, the two lowest hyperfine ground states of fermionic
$^6$Li atoms in a simple cubic optical lattice were labeled as spin-up and spin-down
states. The repulsive interaction between atoms in these states was controlled by a magnetic
Feshbach resonance. Since the magnetic superexchange interaction is given by $J=4t^2/U$, the experiment
controlled the value of $J$ and in a particular regime, it measured antiferromagnetic correlations as
extracted from the spin structure factor. Very recently it was demonstrated that spin (and charge)
correlations can be detected also with the help of site-resolved imaging. In Refs.~\cite{parsons2015,cheuk2016,mazurenko2016} quantum gas microscopy was used to determine spatial correlations for fermionic atoms in a two-dimensional optical lattice. While there is no phase transition in 2D, the measurements have shown an increase of the correlation length as the temperature was lowered. Similar antiferromagnetic correlations extending up to three lattice sites have also been observed in a 1D system \cite{boll2016}.
\\
\\
The simplest many-body model that has a nonzero phase transition in two dimensions is the Ising model~\cite{ising}. Its fermionic analog, the Falicov-Kimball model~\cite{falicov_kimball}, also displays a nonzero transition temperature in two dimensions, which behaves Ising-like when the interaction strength becomes large. This system can be easily simulated with mixtures of cold atoms on optical lattices, because it involves mobile fermions interacting with localized fermions~\cite{ates_ziegler}. One simply needs to have the hopping of the two atomic species to be drastically different. The simplest case of one trapped atomic state for each of the fermionic species maps onto the spinless version of the Falicov-Kimball model. This model has been solved exactly in infinite dimensions via dynamical mean-field theory (DMFT)~\cite{brandt_mielsch,freericks_review} and numerically in two dimensions with Monte Carlo (MC)~\cite{maska_qmc}.

Our motivation for this work stems from the DMFT solution to the problem. There, one can derive a condition for the transition to an ordered phase with a checkerboard pattern~\cite{brandt_mielsch,freericks_review}, which takes the form $1=\sum_n\gamma(n)$, with the sum running over all integers [which label fermionic Matsubara frequencies $i\omega_n=\pi i (2n+1)T$, with $T$ the temperature]. The function $\gamma(n)$ is a complicated function that is constructed from the 
mobile fermion Green's function, its self-energy, the on-site interation between the localized and mobile fermions $U$ and the density of the localized fermions $w_1$. The important point to note, is that if we increase the degeneracy of the mobile fermions (while enforcing that they do not interact with themselves), then the $T_c$ equation is modified by $\gamma(n)\rightarrow N\gamma(n)$, where $N$ is the number of degenerate states for the mobile fermions~\cite{freericks_review}. Since one can immediately show that $\sum_n\gamma(n)\rightarrow C/T$ for $T\rightarrow 0$ and $\sum_n\gamma(n)\rightarrow C'/T^4$ for $T\rightarrow\infty$~\cite{brandt_mielsch}, we expect that the transition temperature for the degenerate system will initially grow linearly in $N$ and then turn over to a slower increase, proportional to $N^{1/4}$ for larger $N$. It is the rapid growth with degeneracy for small $N$, which makes these effects so spectacular. (These ideas are further supported by the observation that increasing species degeneracy {\it lowers} the final temperature after the optical lattice is ramped up in alkaline-earth systems~\cite{hazzard})

The argument that $T_c$ grows linearly with the degeneracy at low temperature can be made more general. We start with the Hamiltonian for the Falicov-Kimball model on a lattice $\Lambda$ that has $|\Lambda|$ lattice sites.
The Hamiltonian for a given configuration of the heavy atoms $\{w\}$ is
\begin{eqnarray}
  \mathcal{H}(\{w\})&=&-t\sum_{\langle ij\rangle}\sum_{\sigma=1}^Nc^\dagger_{i\sigma}c_{i\sigma}
  +U\sum_i\sum_{\sigma=1}^N n_{i\sigma}w_i\nonumber\\
&=&\sum_{\sigma=1}^N \mathcal{H}_\sigma(\{w\}),
\label{eq:hamiltonian}
\end{eqnarray}
where $\sigma$ denotes the $N$ different ``flavors'' of the mobile fermions and $w_i=1$ or 0, denotes whether site $i$ has a localized fermion on it, or not, respectively (the localized fermions continue to be spinless). The hopping matrix is chosen to be nonzero only for nearest neighbors, and we set $t=1$ as our energy unit (we also set $k_B=1$). We define $E_i\equiv \varepsilon_i-\mu$, with $\mu$ the chemical potential and $\{\varepsilon_i\}$ the set of
(degenerate) eigenvalues of $\mathcal{H}_\sigma(\{w\})$, which is independent of the specific value of $\sigma$ because the mobile fermions are noninteracting amongst themselves, and they share the same interaction with the localized fermions. Here, the index $i$ runs over $i=1,\ldots,|\Lambda|$
(we will be working on a square lattice of edge $L$ which then has $|\Lambda|=L\times L$). 

The corresponding grand partition function is given by
\begin{equation}
  \mathcal{Z}=\sum_{\{w\}} \prod_{i=1}^{|\Lambda|}\left[1+e^{-\beta E_i(\{w\})}\right]^N,
  \label{eq:z}
\end{equation}
with $\beta=1/T$ the inverse temperature.
Introducing the free energy ${\cal F}$, Eq. (\ref{eq:z}) can be rewritten as
\begin{equation}
  \mathcal{Z}=\sum_{\{w\}} e^{-\beta{\cal F} (\{w\})},
  \label{eq:z1}
  \end{equation}
where
\begin{eqnarray}
  {\cal F}(\{w\})&=&-\frac{N}{\beta}\sum_i\ln\left[1+e^{-\beta E_i(\{w\})}\right] \nonumber \\
&=& N\sum_i E_i\theta\left[-E_i(\{w\})\right]\nonumber\\
&-& \frac{N}{\beta}\sum_i \ln\left[1+e^{-\beta \left|E_i(\{w\})\right|}\right],
  \label{eq:free_en}
\end{eqnarray}
and $\theta(\ldots)$ is the Heaviside unit step function. In the low-temperature limit the second term on the RHS vanishes.
Inserting the limiting form of ${\cal F}$ into Eq.~(\ref{eq:z1}) yields
\begin{equation}
  \mathcal{Z}=\sum_{\{w\}}e^{-{\beta N} \sum_i E_i\theta\left[-E_i(\{w\})\right]}.
  \label{eq:z2}
\end{equation}
Note, that this result can be recognized to be the condition for the filling of mobile fermions into the Fermi sea determined by the bandstructure corresponding to the particular configuration of the localized fermions, as given by the configuration $\{w\}$.
Since, in the low-temperature limit ${\cal F}$ does not depend
on temperature, the partition function depends on temperature only through the term
$\beta N$. This means that the thermodynamics of the system depends only on the ratio
$T/N$, with initial corrections expected to be small as $T$ rises (because they will be proportional to $T/T_F$ with some suitably large Fermi temperature $T_F$). As a result the critical temperature $T_c$ in the low-temperature limit will necessarily
increase linearly with increasing degeneracy $N$. This is an exact result, independent of the details of the lattice or the dimensionality---it only requires there to be a phase transition.

\begin{figure}
    \centering
    \includegraphics[width=0.48\textwidth]{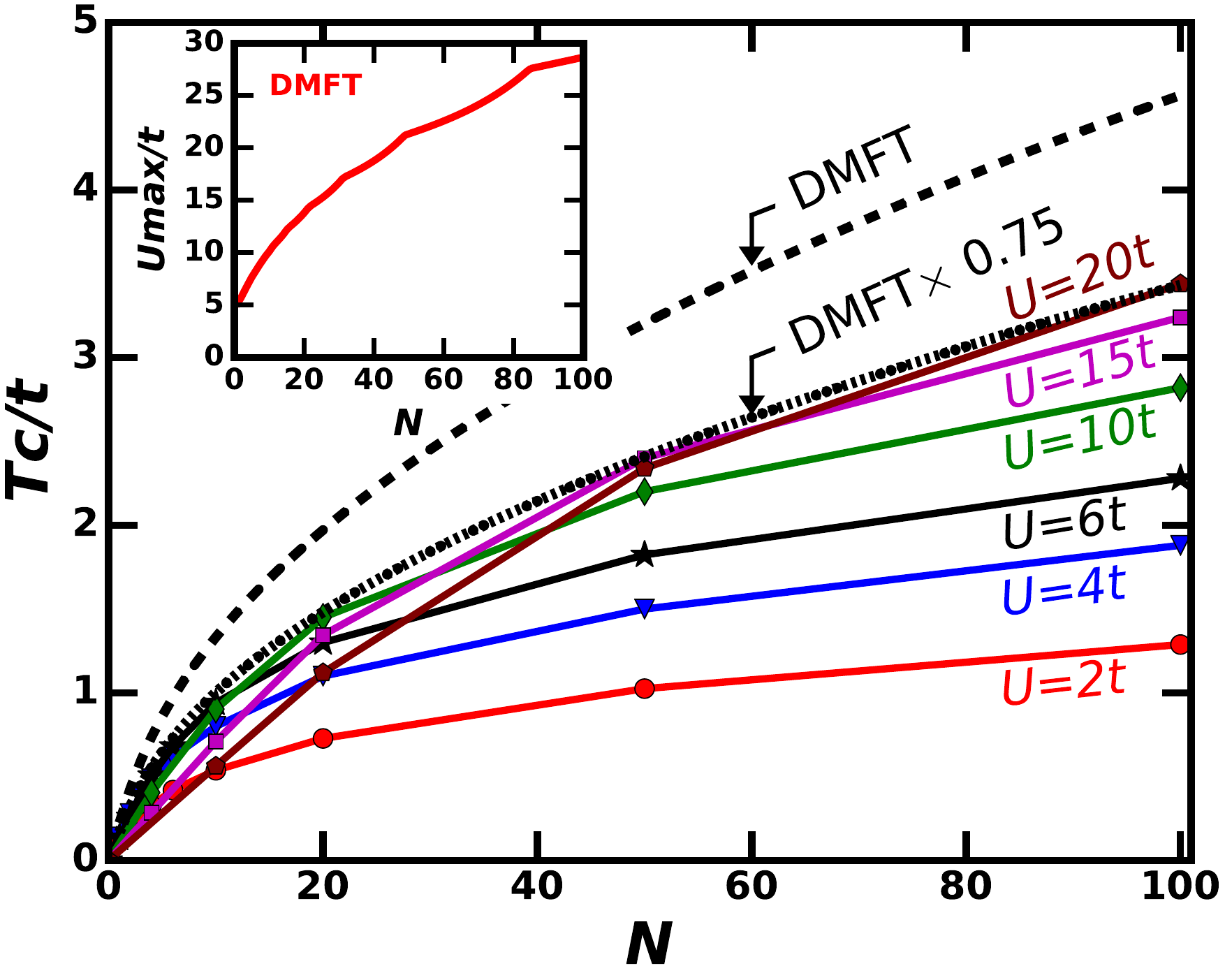}
    \caption{(Color on-line.) The maximal critical temperature $T_c$ plotted as a function of mobile fermion degeneracy $N$ (as calculated with MC). The
      dashed line shows the corresponding DMFT $T_c$ calculated at $U=U_{\rm max}({\rm DMFT})$. The solid lines
      show MC $T_c$'s for different values of $U$. The black dotted line shows $T_c({\rm DMFT})\times 0.75$, which agrees well with nearly all the MC results. In the inset,
      $U_{\rm max}({\rm DMFT})$ is plotted as a function of degeneracy $N$, indicating it changes significantly with $N$.}
    \label{fig:Tc_vs_NU}
\end{figure}

There are two assumptions that went into this anaylsis, which turn out not to hold when we actually calculate the maximal $T_c$ as a function of $N$. First, the lowest $T_c$ values are not so low, so the linear regime fairly rapidly crosses over to a slower increasing behavior and second, the interaction value $U_{\rm max}(N)$, where the maximal $T_{c,{\rm max}}(N)$ occurs, actually changes with $N$ (see the inset in Fig.~\ref{fig:Tc_vs_NU}), so the arguments about the precise functional dependence of the $T_{c,{\rm max}}(N)$ on $N$ turns out not to hold in the actual data; our arguments assumed we compared systems with the same $U$. The first effect is to reduce how $T_c$ increases with $N$, while the second enhances how $T_c$ increases with $N$.

Corrections to the linear dependence of $T_c$ on $N$ come mostly from states close to the Fermi level [$E_i\approx 0$, see the second term in the RHS of Eq. (\ref{eq:free_en})]. Therefore, we can expect that the linear section of the $T_c(N)$ curve can be longer for bipartite lattices for which the density of states is reduced close to the Fermi energy, e.g., for a hexagonal lattice. Also in 3D, where there is no Van Hove singularity (the singularity for the square lattice is reduced by the interaction with the heavy atoms) the linear part can persist to even higher temperatures.

\begin{figure}
    \centering
    \includegraphics[width=0.48\textwidth]{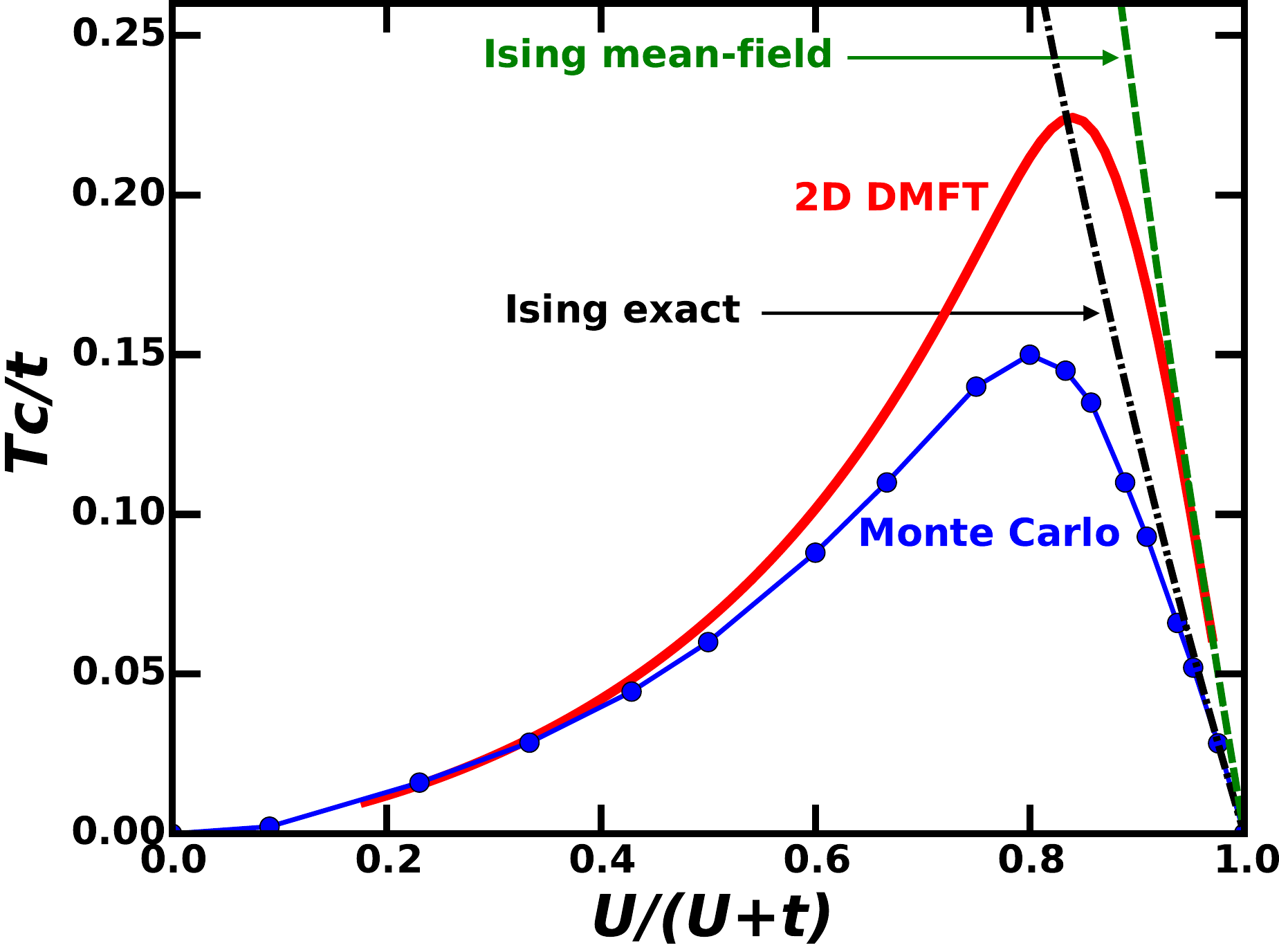}
    \caption{(Color on-line) Comparison of the 2D DMFT (solid red line) and MC (blue dots, connected with a solid line as a guide to the eye) critical temperatures to the checkerboard density wave at half filling for both species on a square lattice with $N=1$. The lines marked
      as "Ising mean-field" (green) and "Ising exact" (black) show the critical temperatures for the corresponding
      Ising model, which become exact for the respective theories when $U\rightarrow\infty$.}
    \label{fig:N=1}
\end{figure}

In Fig.~\ref{fig:N=1}, we plot the transition temperature to the checkerboard density wave on a square lattice with $N=1$. The top curve is for the DMFT approximation, while the bottom curve is for the exact MC results. Note that the interaction strength for the peak of the curve lies in the range of $U\approx 4-5$ with the maximal $U$ value slightly higher for DMFT versus MC. The DMFT results are semiquantitative, and clearly overestimate the $T_c$, but the overall error is not that large.

As $N$ increases, we find that the maximum $T_c$ increases as does the value of the interaction strength where the $T_c(U)$ curve is maximized. The full curve out to $N=100$ is plotted in Fig.~\ref{fig:Tc_vs_NU}. The DMFT results are calculated for each $N$ by first finding the interaction strength at the maximum of the $T_c$ curve. For the MC results, we work with fixed $U$, varying $N$ and then constructing the ``maximal hull'' of the data. It turns out that these MC results are nearly perfectly fit to the DMFT results when the latter are renormalized by a factor of 0.75. The DMFT curve initially grows linearly with $N$, but then settles into an increase that grows proportional to $\sqrt{N-1.7}$, which is in between our linear and 0.25 power results, as we expected, due to the fact that $U_{\rm max}$ increases with $N$.

We find the enhancement of the maximal $T_c$ for higher $N$ versus $N=1$, given by $T_{c,{\rm max}}(N)/T_{c,{\rm max}}(1)$ satisfies: 1.98 (MC, $N=2$), 1.899 (DMFT, N=2); 2.84 (MC, $N=3$), 2.651 (DMFT, $N=3$); and 3.60 (MC, $N=4$), 3.287 (DMFT, $N=4$).
Since the maximal $T_c({\rm DMFT})$ for the Falicov-Kimball model is about one half the maximal $T_c({\rm DMFT})$ for the corresponding Hubbard model, we need to be able to have a degeneracy of $N\ge 3$ before this effect will have a high enough $T_c$ that it can reach current experimentally accessible values for the 3D case. We focus the remainder of this letter on discussing possible experimental realizations for such higher degeneracy mixtures.

The Falicov-Kimball model has zero interaction between the mobile fermions. One can argue, on rather general grounds, that the modification of $T_c$ due to a nonzero intraspecies interaction $u$ will have corrections to $T_c$ of order $u^2$. Hence, if $u$ is small, the effect we discuss here should continue to hold, with only slight reductions. This allows us to formulate our search criterion for physical systems that will show this degenerate species effect.

In searching for appropriate mixtures, we want to find systems that (i) can have a degeneracy of three or more for the light fermionic species, (ii) have a similar interspecies interaction $U$ between the mobile and localized fermions, which will be tuned either via an interspecies Feshbach resonance, or via the depth of the trapping potential for the light species; and (iii) have a small intraspecies interaction $u$ between the mobile fermions. We also note, that as long as the localized particle is nondegenerate, then it can actually be either Bose or Fermi, since its statistics does not enter the analysis because it does not move. (However, if the heavy particle is a boson, we do need its intraspecies interaction to be large and positive, so it generically forms a Mott insulator with at most one particle per site and it does not Bose condense on the lattice.)

We start with examining some prototypical systems which have already been demonstrated to be trapped on optical lattices. The first choice to examine is mixtures of $^{40}$K (mobile fermion) and $^{87}$Rb (localized boson)~\cite{krb}. If we could trap the $m_F=-5/2,-7/2$, and $-9/2$ states of K, we would have an $N=3$ mixture. This system is nice, in the sense that it has a tunable interspecies interaction via a Feshbach resonance, and the intraspecies interactions for K have a scattering length on the order of  $100~a_0$ (in some cases one of the pairs can be tuned to zero scattering length). The challenge is that the Rb-Rb interaction is too small (on the order of $100~a_0$), and is not tunable, which would make it difficult to satisfy the required conditions for this effect. If we instead try $^{133}$Cs (localized boson)~\cite{KCs}, we find that the Cs-Cs interaction is large, with a scattering length near $2000~a_0$ at $B\approx 260$~G, but the interspecies interaction is small  ($\approx -40~a_0$) and not simultaneously tunable for all three K species.

Moving on to other possibilities, if we use mixtures of $^{171}$Yb or $^{173}$Yb (mobile fermion)~\cite{Yb1,Yb2} and $^{133}$Cs (localized boson)~\cite{YbCs1,YbCs2}, we only have a degeneracy of $N=2$ for $^{171}$Yb, even though its intraspecies scattering is small, while for $^{173}$Yb the intraspecies scattering length is $\approx 200~a_0$, which is still viable, given the potentially large Cs-Cs scattering length, but it would require a tunable Cs-Yb scattering length that is large, and although this has not yet been measured, we do not anticipate that there is any reason why it should be particularly large. If we tried Rb as the localized boson~\cite{YbRb}, it suffers from the same issues as with K-Rb{\textemdash}namely, the Rb-Rb scattering is too small.

Using $^6$Li as the mobile fermion appears attractive~\cite{Li1,Li2}. However, the interspecies scattering length is only small for low fields, and when a mixture is formed from the $N=3$ trappable state,  at least one intraspecies interaction will be large (although the other two can be close to zero). So, this case is suboptimal.

Next, we consider mixtures of $^{87}$Sr (light fermion) which has up to $N=10$ and a Sr-Sr scattering length on the order of $100~a_0$~\cite{Sr1,Sr2}. If we use Cs as the (localized boson), then if the Cs-Cs scattering length can be set to the order of a few $1000~a_0$, and the Sr-Cs scattering length is on the order of $500~a_0$, then this system might work to illustrate this degenerate species effect, and it has the potential to be spectacularly large.

The remaining choices that might be workable seem to be longshots, but cannot yet be ruled out because we do not have enough information about their interspecies interactions. We discuss some of these possibilities next. 

$^{43}$Ca is a fermion with a nuclear spin of 7/2~\cite{Ca1,Ca2}, $^{25}$Mg is a fermion with a nuclear spin of 5/2~\cite{Mg}, Ba has two spin 3/2 fermionic species~\cite{Ba}, and $^{201}$Hg is also spin 3/2~\cite{Hg}. It is unknown what the intraspecies interactions are amongst these different spin states, how many can be trapped, and what their interspecies interactions are with potential heavy particles. So they all are possible, but at this stage quite difficult systems to work with. Finally, there are all of the magnetic-dipole systems, like Er~\cite{Er1,Er2}, Dy~\cite{Dy1,Dy2}, and Cr~\cite{Cr1,Cr2,Cr3}. These systems often have chaotic intraspecies interactions due to a huge number of resonances, but they might show some small interactions at low fields, and hence may also be viable candidates for the light fermions.

In summary, we have illustrated the idea that by enhancing species degeneracy, one can enhance $T_c$ for fermionic neutral atoms trapped on optical lattices such that their $T_c$ to an ordered state can be raised high enough that they would be accessible to explore with current experimental technology in cooling. This idea comes at this problem from a different angle than the many different cooling strategies that have been proposed, and could provide the ability to truly study spatially ordered quantum phases. The challenge is to find the right mixture of atoms where this effect can be fully exploited. We have suggested some possible systems, with Yb-Cs and Sr-Cs mixtures as the most promising, but it is clear the experiments will be challenging to carry out. We want to end by commenting that similar work has examined $SU(N)$ symmetric Hubbard models. The repulsive case actually sees a decrease in the antiferromagnetic $T_c$ with increasing $N$~\cite{japanese1}, while the attractive case sees an enhancement similar to what we see for the density-wave instability~\cite{japanese2}, we do not know of any large $N>3$ systems with attractive interactions. Furthermore, there are challenges with finding atomic systems with a small enough $U$ value (for large  $N$), since a maximal hopping is required to have an accurate single-band description.

K.N. acknowledges support from the National Science Foundation under grant number PHY-1314295.
M.M.M. acknowledges support by National Science Centre (NCN) under Grant No. DEC-2013/11/B/ST3/00824.
K.D. acknowledges support from the National Science Foundation under grant number DMR-1358978.
J.K.F. acknowledges support from the National Science Foundation under grants numbered PHY-1314295 for the initial stage of the work and PHY-1620555 for the final stage. He was also supported by the McDevitt bequest at Georgetown University.

\end{document}